\documentclass[12pt]{article}
\usepackage{epsfig}
\usepackage{amssymb}
\usepackage{amsmath}
\usepackage{amsfonts}

\newcommand{\R}{\mathbb{R}}

\newcommand{\be}{\begin{equation}}
\newcommand{\ee}{\end{equation}}
\newcommand{\bea}{\begin{eqnarray}}
\newcommand{\eea}{\end{eqnarray}}

\newcommand{\kt}{\rangle}
\newcommand{\br}{\langle}

\newcommand{\ed}{\end{document}}

\begin{document}

\title{Comment on ``{\cal PT}-Symmetric versus Hermitian Formulations
of Quantum Mechanics''}

\author{\\
Ali Mostafazadeh
\\
\\
Department of Mathematics, Ko\c{c} University,\\
34450 Sariyer, Istanbul, Turkey\\ amostafazadeh@ku.edu.tr}
\date{ }
\maketitle

\begin{abstract}
We explain why the main conclusion of Bender et al, J.~Phys.~A
{\bf 39}, 1657 (2006), regarding the practical superiority of the
non-Hermitian description of ${\cal PT}$-symmetric quantum systems
over their Hermitian description is not valid. Recalling the
essential role played by the Hermitian description in the
characterization and interpretation of the physical observables,
we maintain that as far as the physical aspects of the theory are
concerned the Hermitian description is not only unavoidable but
also indispensable.

\vspace{5mm}

\vspace{2mm}


\end{abstract}






Recently Bender, Chen, and Milton \cite{bcm} have employed the
path integral method to examine the perturbative calculation of
the ground-state energy and a one-point Green's function for the
${\cal PT}$-symmetric cubic anharmonic oscillator. They performed
this calculation in both the ${\cal PT}$-symmetric
(pseudo-Hermitian) and the Hermitian descriptions of this model
and concluded that the use of the latter description leads to
practical difficulties that are ``severe and virtually
insurmountable $\cdots$'', and that such difficulties do not arise
in the former description.

As explained in \cite{cjp-2004b}, the level of the difficulty of a
calculation in ${\cal PT}$-symmetric quantum mechanics depends on
the quantity that one chooses to calculate. If one wishes to
calculate the expectation value of a canonical pair of basic
observables of the theory, both the descriptions/representations
involve dealing with practical problems with the same degree of
difficulty. The reason why the difficulties with the ${\cal
PT}$-symmetric representation do not surface in the interesting
calculations reported in \cite{bcm} is that the authors only
calculate the ground-state energy and the one-point Green's
function for the operator $x$.

The difficulties with the calculation of ground- and excited-state
energies in the Hermitian representation is already clear from the
complicated expression for the Hermitian Hamiltonian that was
obtained in \cite{jpa-2005b,jones-jpa-2005}, and indeed the ${\cal
PT}$-symmetric representation is much more convenient for this
purpose whether one uses the path-integral or operator methods.
The situation for the calculation of the expectation value of the
observables such as the position which may be related to the
corresponding Green's functions must be treated with more care,
for the very notion of the observable for a ${\cal PT}$-symmetric
system is a delicate issue
\cite{critique,comment,cjp-2004b,jpa-2005b}. Specifically, for
${\cal PT}$-symmetric models such as the one considered in
\cite{bcm}, the $x$ operator is not an observable. The same holds
for the unitary-equivalent operator $\tilde x=e^{-Q/2}xe^{Q/2}$
introduced in \cite{bcm} which is manifestly non-Hermitian and
consequently fails to be an observable in the Hermitian
representation (conventional quantum mechanics.)

As described in \cite{critique,jpa-2004b}, the position observable
is the nonlocal pseudo-Hermitian operator $X=e^{Q/2}xe^{-Q/2}$,
and the physical one-point Green's function is $\br
0|X|0\kt_{_{\cal CPT}}$. The one-point Green's function $\br
0|x|0\kt_{_{\cal CPT}}$ calculated in \cite{bcm}, which takes an
imaginary value, does not represent a physical quantity.

It is not difficult to see that the calculation of the physical
one-point Green's function $\br 0|X|0\kt_{_{\cal CPT}}$ is as
difficult in the ${\cal PT}$-symmetric representation of the model
as the calculation of the non-physical one-point Green's function
$\br 0|x|0\kt_{_{\cal CPT}}$ is in the Hermitian representation.
Therefore as far as the calculation of physical quantities such as
$\br 0|X|0\kt_{_{\cal CPT}}$ are concerned the ${\cal
PT}$-symmetric description is not superior to the Hermitian
description.

The key ingredient that makes the Hermitian description of ${\cal
PT}$-symmetric systems indispensable is its crucial role in
determining the operators that represent the physical observables
of the theory as well as its utility in providing a physical
interpretation for these operators \cite{cjp-2004b,jpa-2004b}. In
particular, once the associated Hermitian Hamiltonian is
determined one can identify the underlying classical system and
assign physical meaning to the Hamiltonian using its classical
counterpart. This has been achieved for the ${\cal PT}$-symmetric
cubic anharmonic oscillator in \cite{jpa-2005b} and for some other
toy models in \cite{jpa-2004b,jmp-2005,banerjee,p67}. For example,
it was an examination of the Hermitian representation of the
${\cal PT}$-symmetric cubic anharmonic oscillator that revealed
the curious fact that up to cubic terms in the perturbation theory
it described a Hermitian quartic anharmonic oscillator with a
certain position-dependent mass \cite{jpa-2005b}. See also
\cite{banerjee,p67}. This has recently motivated a study of
position-dependent mass Hermitian Hamiltonians in terms of their
constant-mass non-Hermitian equivalent Hamiltonians
\cite{bagchi-q-r}. Other evidences for the importance of the
Hermitian representation is its role in the proof of the physical
equivalence of the ${\cal PT}$-symmetric and conventional massive
Thirring and Sine-Gordon field theories \cite{bjr} and the recent
construction of the equivalent Hermitian Hamiltonian for the
${\cal PT}$-symmetric Hamiltonian $H=p^2-x^4$ defined on a complex
contour \cite{jones-2006}. The latter is particularly significant,
for the equivalent Hermitian Hamiltonian turns out to be
$h=16p^2+x^4/64-x/2$ that is defined on $\R$. This observation not
only explains the physical meaning of $H$ but it also provides a
direct proof of the reality, positive-definiteness, and
discreteness of its spectrum, a fact whose initial proof required
quite sophisticated mathematical tools \cite{dorey}.

In conclusion, we maintain that the importance of the Hermitian
representation of the ${\cal PT}$-symmetric (and other) unitary
quantum systems cannot be undermined by a demonstration that
certain calculations take a simpler form in the non-Hermitian
representation of these systems. There are other physically
relevant calculations that are at least as difficult in the
non-Hermitian representation as they are in the Hermitian
representation. More importantly, the quest for understanding the
physical meaning and potential applications of such systems makes
the Hermitian representation absolutely essential. Finally, we
wish to emphasize a positive aspect of the results of \cite{bcm}
namely the curious identities between sums of certain Feynman
diagrams that follow from the unitary-equivalence of the two
representations.

{

}

\ed